\documentclass[doublecol]{epl2} 
\usepackage{amsmath,amssymb}
\usepackage{stkernel}
\usepackage{flushend}
\newcommand{\rd}{\mathrm{d}}
\title{Elastic turbulence in a shell model of polymer solution}

\author{Samriddhi Sankar Ray\inst{1} \and Dario Vincenzi\inst{2}}
\shortauthor{Samriddhi Sankar Ray and Dario Vincenzi}

\institute{                    
  \inst{1} International Centre for Theoretical Sciences, Tata Institute of 
Fundamental Research - Bangalore 560089, India\\
  \inst{2} Laboratoire J. A. Dieudonn\'e, Universit\'e Nice Sophia Antipolis,
CNRS, UMR 7351 - Nice 06108, France
}

\abstract{
We show that, at low inertia and large elasticity, 
shell models of viscoelastic fluids
develop a chaotic 
behaviour with properties similar to those of elastic turbulence. 
The low dimensionality of shell models allows us to
explore a wide range both 
in polymer concentration and in Weissenberg number. 
Our results demonstrate that the physical mechanisms at the origin of 
elastic turbulence do not rely on the boundary conditions or on the geometry 
of the mean flow. 
}

\begin{document}

\maketitle

\section{Introduction}

Elastic turbulence is a chaotic regime that develops in low-inertia
viscoelastic fluids when the elasticity of the fluid exceeds a critical
value~\cite{GS00}.  It is characterised by power-law velocity spectra (both in
time and in space) and by a strong increase of the flow resistance compared to
the laminar regime.  Elastic turbulence differs from hydrodynamic
turbulence in that inertial nonlinearities are irrelevant and the chaotic
behaviour of the flow is entirely generated by elastic instabilities. In
addition, the spatial spectrum of the velocity decays faster than in
hydrodynamics turbulence; thus, the velocity field is smooth in
space.  Elastic turbulence has important applications, since the
possibility of inducing instabilities at low Reynolds numbers allows the
generation of mixing flows in microfluidics devices~\cite{GS01,BSBGS04}. This
phenomenon has been used, for instance, to study the deformation of DNA
molecules in chaotic flows~\cite{LS14}. Furthermore, elastic turbulence
provides a possible explanation of the improvement in oil-displacement
efficiency that is observed when polymer solutions are used to flood reservoir
rocks~\cite{MLHC16}.

The first experiments on elastic turbulence have used confined flows with
curved stream lines~\cite{GS04}. Nonetheless, purely elastic instabilities have
been shown to develop also in a viscoelastic version of the Kolmogorov flow,
which is periodic and parallel~\cite{BCMPV05,BBCMPV07}. Indeed, elastic turbulence in
the viscoelastic Kolmogorov flow exhibits a phenomenology qualitatively
similar to that observed in experiments~\cite{BBBCM08,BB10}.
Low-Reynolds-number elastic instabilities have also been predicted for the
Poiseuille flow~\cite{MSMS04}  and for the planar Couette flow~\cite{MS05} of a
polymer solution at large elasticity.  More recently, elastic turbulence has
been observed experimentally in a straight microchannel~\cite{PMWA13,BBMMKC15}
and numerically in a periodic square~\cite{G14}.  These findings indicate that
elastic turbulence also develops in simplified flow configurations and that the
specific geometry of the system may not play a crucial role in this phenomenon.
In this Letter, we take a step further in this direction and study elastic
turbulence in a shell model of polymer solution.

Hydrodynamical shell models are low-dimensional models that preserve the
essential shell-to-shell energy transfer feature of the original partial
differential equations in Fourier space.  Despite the fact that they are not
derived from the principle hydrodynamic equations in any rigorous way, they
have played a fundamental role in the study of fluid turbulence since they are
numerically tractable~\cite{F95,BJPV98,Biferale,PPR09}.  Shell models have also
achieved remarkable success in problems related to passive-scalar
turbulence~\cite{JPV92,WB96,MP05,RMP08}, magnetohydrodynamic
turbulence~\cite{PSF13}, rotating turbulence~\cite{Sagar}, binary
fluids~\cite{JO98,RB11}, and fluids with polymer
additives~\cite{BdAGP03,KGP05}.  Furthermore, the mathematical study of shell
models has yielded several rigorous results, whose analogs are still lacking
for the three-dimensional Navier--Stokes equations (e.g.,
Refs.~\cite{CLT06,F11}).

A shell model of polymer solution can be obtained by coupling the evolution of
the velocity variables with the evolution of an additional set of variables
representing the polymer end-to-end separation field.  Shell models of polymer
solutions have been successfully applied to the study of drag reduction in
forced~\cite{BdAGP03,BCHP04,BCP04} and decaying~\cite{KGP05} turbulence,
two-dimensional turbulence with polymer additives~\cite{BHP04}, and turbulent
thermal convection in viscoelastic fluids~\cite{BCdA10,BCW14}.  Here, we study
a shell model of polymer solution in the regime of low inertia and high
elasticity.  We show that this shell model undergoes a transition froam a laminar to a 
chaotic regime with properties remarkably similar to those of elastic turbulence.
Moreover, the use of a low-dimensional model allows us to explore the
properties of elastic turbulence over a wide range both in polymer
concentration and in Weissenberg number, which would be difficult to cover with
direct numerical simulations of constitutive models of viscoelastic fluids.

\begin{table}
\centering
{\begin{tabular}{|c|c|c|c|c|c|c|c|c|c|}
\hline
Type & $\epsilon$ & $N$ & $\nu$ & $\delta t$ & $f_0$\\
\hline
\hline
chaotic for $c=0$  & 0.5 & 15 & $10^{-1}$ & $10^{-4}$ & 0.01 \\
\hline
non-chaotic for $c=0$ & 0.3 & 22 & $10^{-6}$ & $10^{-4}$ & 0.01 \\
\hline
\end{tabular}}
\caption{The parameters, defined in the text, for the different sets of 
our simulations.}
\label{dec_para}
\end{table}

\section{Shell model of polymer solution}

We consider the shell model of polymer solution introduced by Kalelkar
\textit{et al.}~\cite{KGP05}, which is based on a shell model initially
proposed for three-dimensional magnetohydrodynamics~\cite{FS98,BSDR98} and
reduces to the GOY model~\cite{G73,YO87} when polymers are absent.  The shell
model by Kalelkar \textit{et al.}~\cite{KGP05} can be regarded as a reduced,
low-dimensional version of the FENE model~\cite{BCAH87}.  It describes the
temporal evolution of a set of complex scalar variables $v_n$ and $b_n$
representing the velocity field and the polymer end-to-end separation field,
respectively.  The variables $v_n$ and $b_n$ evolve according to the following
equations~\cite{KGP05}:

\begin{eqnarray} 
\label{eq:v} 
\frac{\rd v_n}{\rd t}&=&\Phi_{n,vv}-\nu_s k^2_n
v_n+\frac{\nu_p}{\tau_p} P(b)\Phi_{n,bb}+f_n, \\ 
\label{eq:b} 
\frac{\rd b_n}{\rd t}&=&\Phi_{n,vb}+\Phi_{n,bv} -\frac{1}{\tau_p}P(b)b_n-\nu_b k_n^2 b_n.
\end{eqnarray} 
where $n=1,\dots,N$, $k_n=k_0 2^n$, $P(b)=1/(1-\sum_n\vert b_n\vert^2)$ and
$\Phi_{n,vv}=i(a_1k_nv_{n+1}v_{n+2}+a_2k_{n-1}v_{n+1}v_{n-1}+a_3k_{n-2}v_{n-1}
v_{n-2})^*$, $\Phi_{n,bb}=-i(a_1 k_n b_{n+1}b_{n+2}+a_2 k_{n-1}
b_{n+1}b_{n-1}+a_3 k_{n-2} b_{n-1}b_{n-2})^*$, $\Phi_{n,vb}=i(a_4 k_n
v_{n+1}b_{n+2}+a_5 k_{n-1}v_{n-1}b_{n+1}+a_6 k_{n-2} v_{n-1} b_{n-2})^*$, and
$\Phi_{n,bv}=-i(a_4 k_n b_{n+1} v_{n+2}+a_5 k_{n-1} b_{n-1}v_{n+1}+a_6 k_{n-2}
b_{n-1} v_{n-2})^*$ with $k_0=1/16$, $a_1=1$, $a_2=-\epsilon$,
$a_3=-(1-\epsilon)$, $a_4=1/6$, $a_5=1/3$, $a_6=-2/3$, and the single free
parameter $\epsilon$
determines whether or not, in the absence of polymers,
the behaviour of the shell model is chaotic.
The GOY shell model for fluids indeed shows a chaotic behaviour 
for $0.33 \lesssim \epsilon
\lesssim 0.9$ and a non-chaotic behaviour 
for $\epsilon\lesssim 0.33$~\cite{BLLP95,KOJ98}; the standard choice for
hydrodynamic turbulence is $\epsilon=0.5$~\cite{F95,BJPV98,Biferale,PPR09}.
As we shall see later, it is useful to study elastic turbulence in both these regimes.

\begin{figure}[t!]
\centering
\includegraphics[width=\columnwidth]{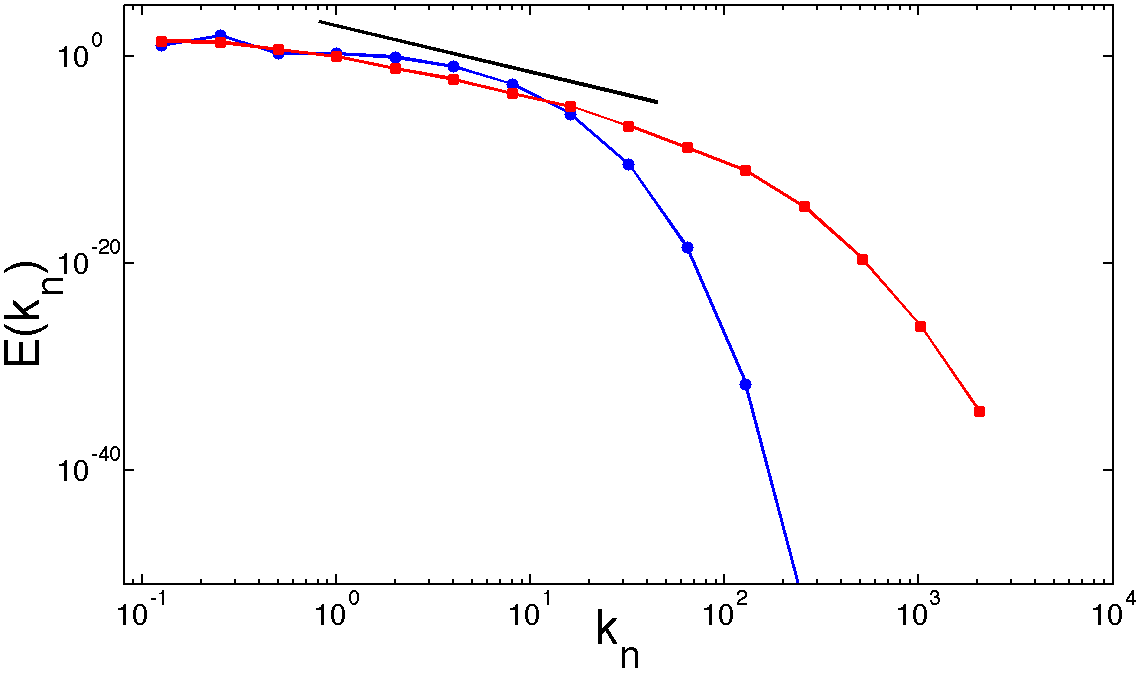}
\caption{Log-log plots of the kinetic energy spectrum $E(k)$ vs the wavenumber $k$
for a highly viscous flow with (red, filled squares) and without (blue, filled
circles) polymer additives (see text). The curve without the addition of
polymers do not show any algebraic scaling. However, the addition of polymers
leads to the development of a power-law scaling $k^{-4}$ in the energy spectrum (as
indicated by the thick black line).}
\label{fig:spectra}
\end{figure}

\begin{figure}
\centering
\includegraphics[width=\columnwidth]{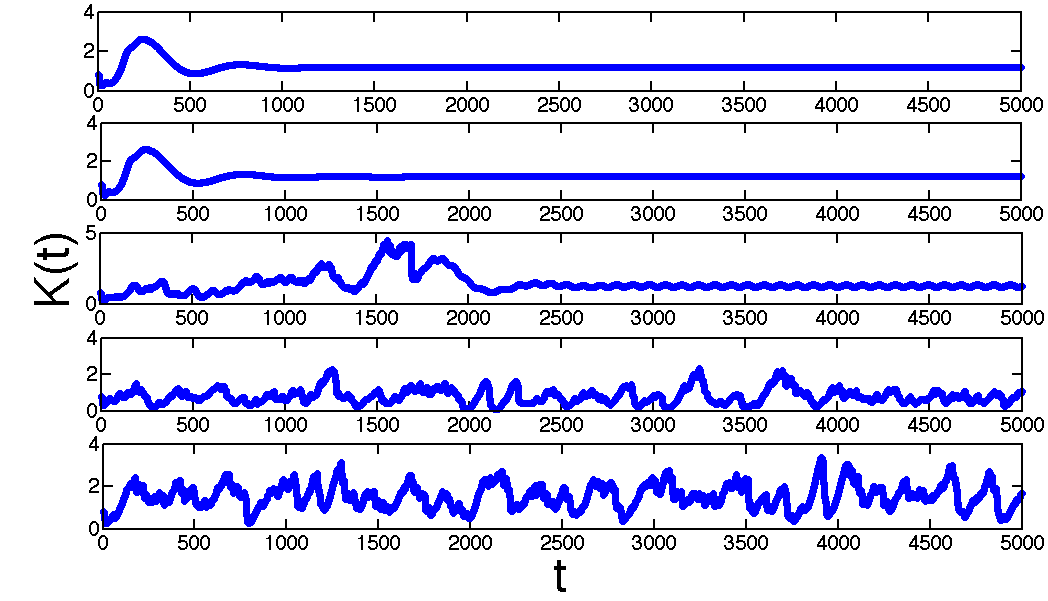}
\caption{The kinetic energy vs time for various values of $c$ and $Wi$. From the uppermost to the
lowermost curve the curves correspond to $c = 1.0$, $Wi \approx 0.25$; $c =
4.0$, $Wi \approx 0.25$; $c = 1.0$, $Wi \approx 25$; $c = 4.0$, $Wi \approx 25$;
and $c = 20.0$, $Wi \approx 25$.  The top two curves show non-chaotic, laminar
behaviour with a transition to periodic dynamics in the middle curve and then fully elastic
turbulence in the bottom two.}
\label{fig:all_energy}
\end{figure}

\dblfloatsatbottom
\begin{figure*}
\centering
\includegraphics[width=0.5\linewidth]{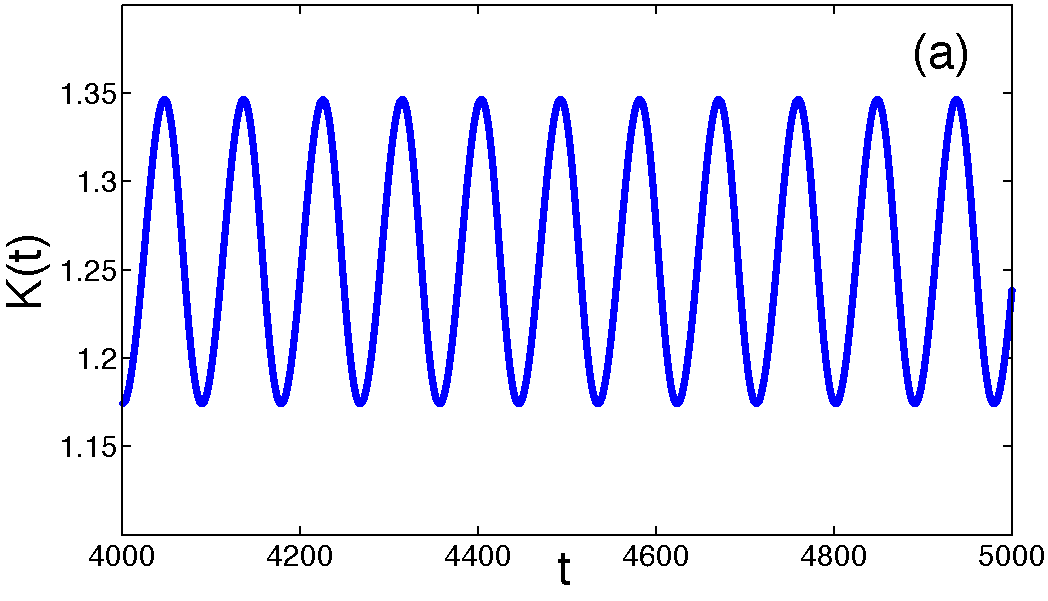}%
\includegraphics[width=0.5\linewidth]{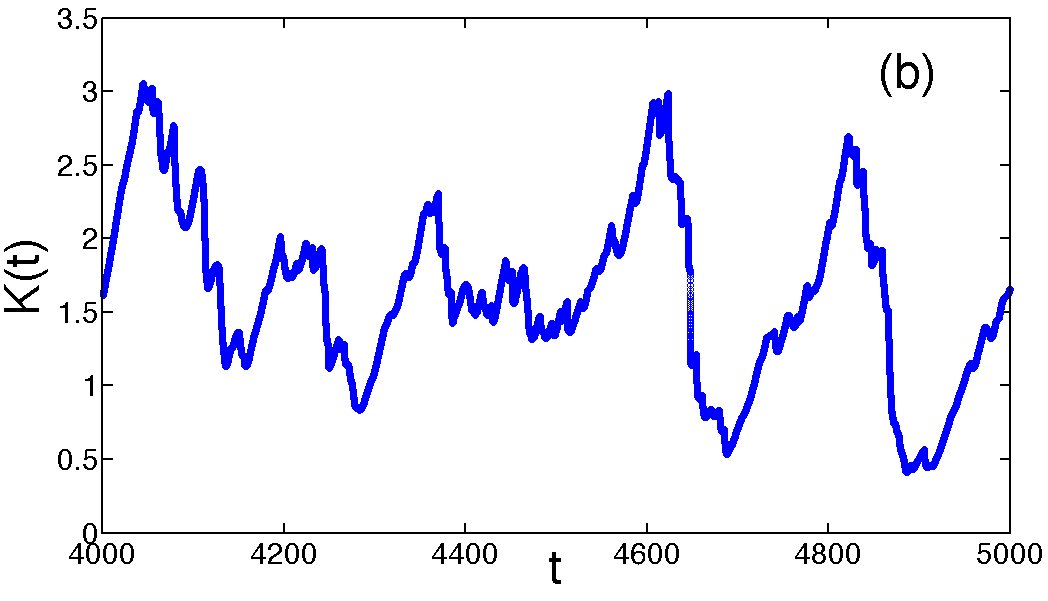}
\caption{The kinetic energy vs time, at $Wi \approx 25$ for (a) $c = 1.0$
and (b) $c = 20.0$ in the time interval
where the flow is statistically stationary.  We see a clear transition from
a periodic behaviour to a fully elastic turbulence regime. 
}
\label{fig:zoomed_energy}
\end{figure*}
\vskip\topskip

The number of shells that are used is given by $N$, the coefficient of
kinematic viscosity by $\nu_s$, the polymer relaxation time by $\tau$, $\nu_p$
is the polymer viscosity parameter, $\nu_b=10^{-13}\nu_s$ is a damping
coefficient to allow for the dissipation term $-\nu_b k_n^2 b_n$ to be added to
Eq.~\eqref{eq:b} in order to improve numerical
stability~\cite{BCHP04,BCP04,KGP05}, and the forcing $f_n$ drives the system to
a non-equilibrium statistically stationary state. In particular we use either a
deterministic forcing $f_n = f_0(1 + \imath)\delta_{n,2}$ or a white-in-time
Gaussian stochastic forcing with amplitude $f_0$ acting on the $n=2$ shell.  We
choose initial conditions of the form $v_n^0 = k_n^{1/2}e^{\imath \phi_n}$ for
$n = 1, 2$ and $v_n^0 = k_n^{1/2}e^{-k_n^2}e^{\imath \phi_n}$ for $3 \le n \le
N$ and, for the polymer field, $b_n^0 = k_n^{1/2}e^{\imath \theta_n}$ for $1
\le n \le N$.  Here $\phi_n$ and $\theta_n$ are random phases uniformly
distributed between 0 and $2\pi$.  Equations~\eqref{eq:v} and~\eqref{eq:b} are
solved numerically through a second-order Adams--Bashforth method with a time
step $\delta t$ for all our simulations. The numerical values for the various
parameters of our simulations are given in Table~1.  


By analogy with continuum models of polymer solutions, we interpret the ratio
$c=\nu_p/\nu_s$ as the polymer concentration~~\cite{BCHP04,BCP04,KGP05}.
We define the mean dissipation rate of the
flow as $\varepsilon=\langle\nu_s\sum_n k_n^2 v_n^2\rangle$; thence we extract
the large-scale time $T=(k_1^2\varepsilon)^{-1/3}$, which allows us to define
the Weissenberg number as $\mathrm{Wi}=\tau/T$. In our simulations we choose
eight different values both of $c$ and of
$\mathrm{Wi}$ such that $0 \le c \le 20$ and $0 \le \mathrm{Wi} \lesssim 25$. 
The Weissenberg number is varied by varying $\tau$, so that the inertia of
the system remains constant and negligible for all Wi.

\section{Elastic-turbulence regime}

In order to understand whether the shell model defined via Eqs.~\eqref{eq:v}
and~\eqref{eq:b} indeed shows the typical features of elastic turbulence, we
perform numerical simulations of the shell model with $\epsilon=0.5$ and a
stochastic forcing.  The parameters  (see Table~\ref{dec_para}) are such that
for $c=0$ the shell model is not turbulent.  The time-averaged (in the steady
state) kinetic-energy spectrum $E(k_n)=\vert v_n\vert^2/k_n$ indeed decreases
sharply with the wavenumber $k_n$ without any apparent power-law scaling (see
the blue line with filled circles in Fig.~\ref{fig:spectra}).%

A typical signature for elastic turbulence is the development of a power-law
energy spectrum with an exponent smaller than $-3$ as the Weissenberg number is
increased at fixed Reynolds number much smaller than~1 \cite{GS04,BSS07}.  We
therefore turn on the polymer field in the shell model ($c \neq 0$), and for
sufficiently large $c$ and $\mathrm{Wi}$ a power-law scaling emerges. 
In  Fig.~\ref{fig:spectra}, the energy spectrum for $c = 20$ and $\mathrm{Wi} =
25$ is shown (red squares).  We see a clear power-law behaviour, namely $E(k_n)
\sim k_n^{-4}$, as is indicated by the thick black line.  The value of the
exponent is close to that found in experiments~\cite{GS04,BSS07} and in
numerical simulations~\cite{BBBCM08,BB10,WG13,WG14} and is consistent with the
theoretical predictions based on the Oldroyd-B model~\cite{FL03}.  The spectrum
of the polymer end-to-end separation field does not show a power-law behaviour
and is concentrated around small wave numbers, in agreement with direct
numerical simulations of elastic turbulence~\cite{G14}.  An analogous behaviour
is found with a deterministic forcing.  This is accompanied by a corresponding
increase of the largest Lyapunov exponent as discussed in detail later.  The
spectrum of the polymer end-to-end separation field does not show a power-law
behaviour and is concentrated around small wave numbers, in agreement with
direct numerical simulations of elastic turbulence~\cite{G14}.  Thus, the shell
model reproduces the most obvious signature of elastic turbulence, namely, the
emergence of a large-scale chaotic dynamics in a {\it laminar} flow with the
addition of polymers. 

\begin{figure*}[t!]
\centering
\includegraphics[width=0.33\linewidth]{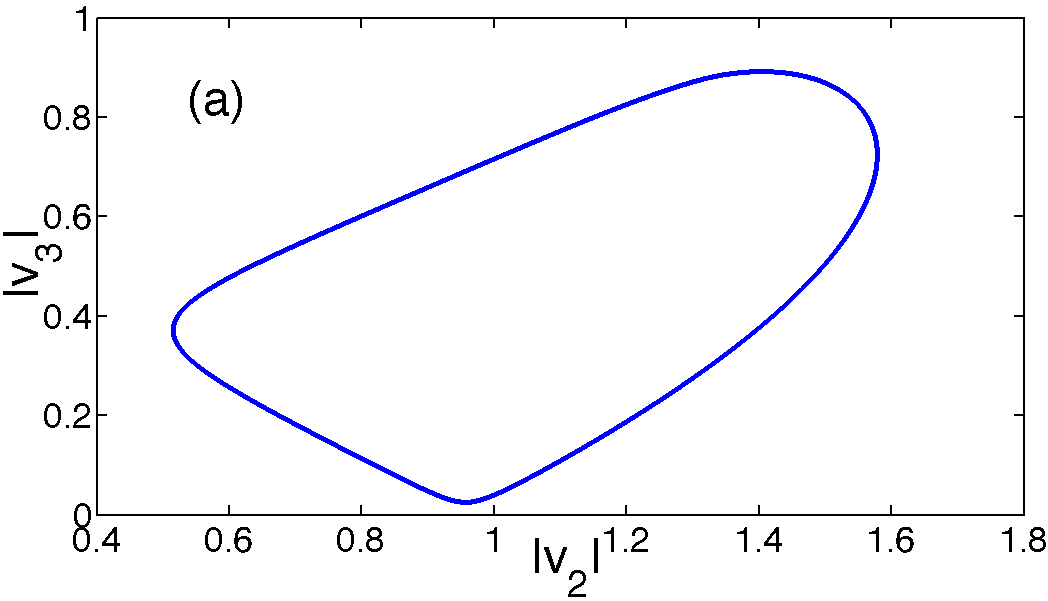}%
\includegraphics[width=0.33\linewidth]{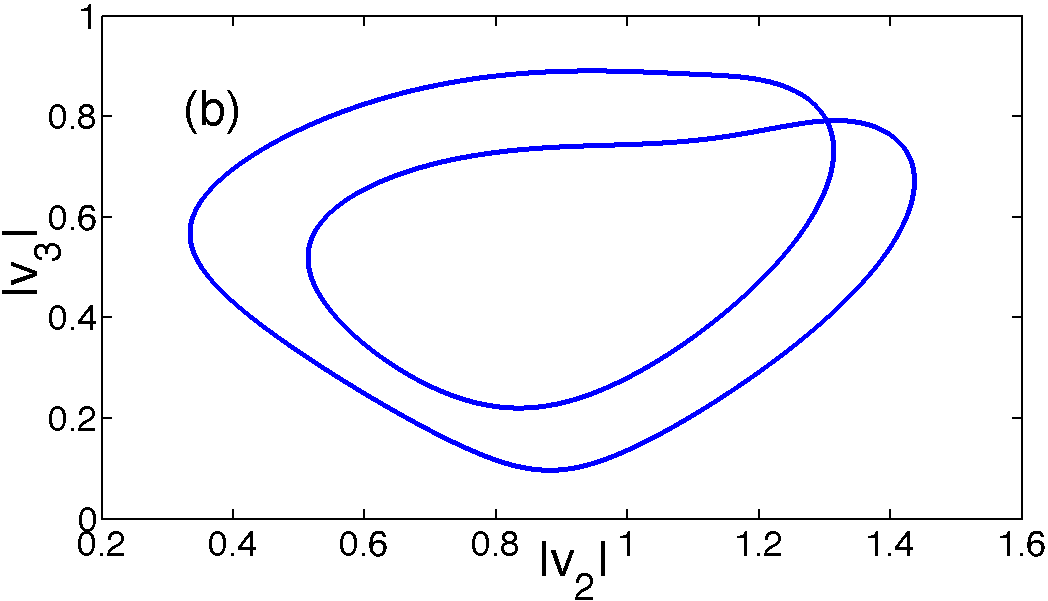}%
\includegraphics[width=0.33\linewidth]{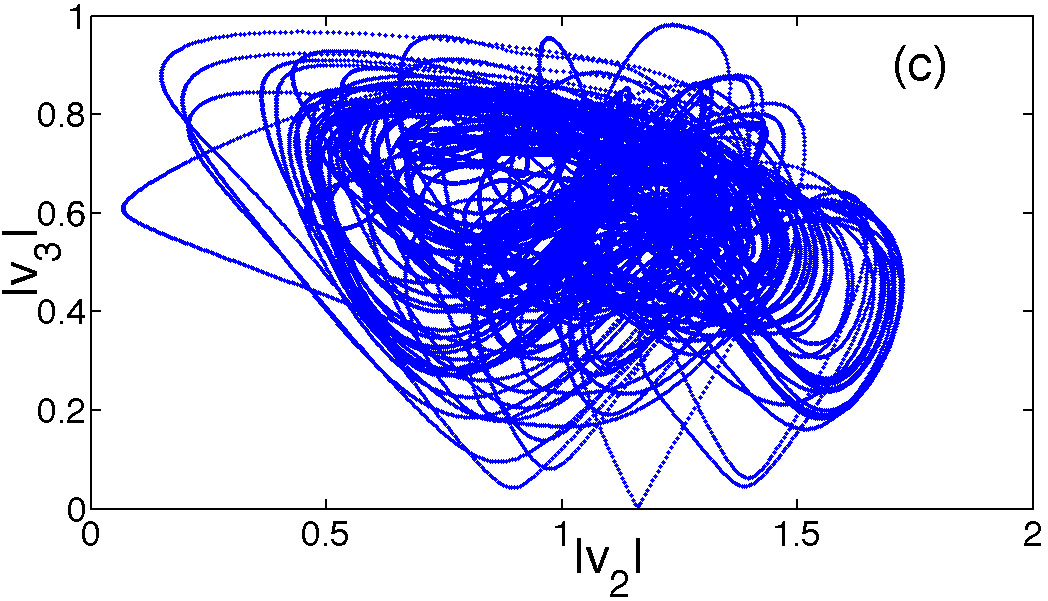}
\caption{The map for $\vert v_{3}\vert$ vs $\vert v_2\vert$, at fixed $Wi = 25$ and (a) $c = 0.5 $ (b) $c = 7.0$ and 
(c) $c = 20.0$.  We see a clear transition from a non-chaotic to a chaotic behaviour.}
\label{fig:map}
\end{figure*}%

\begin{figure*}[t]
\centering
\includegraphics[width=0.45\linewidth]{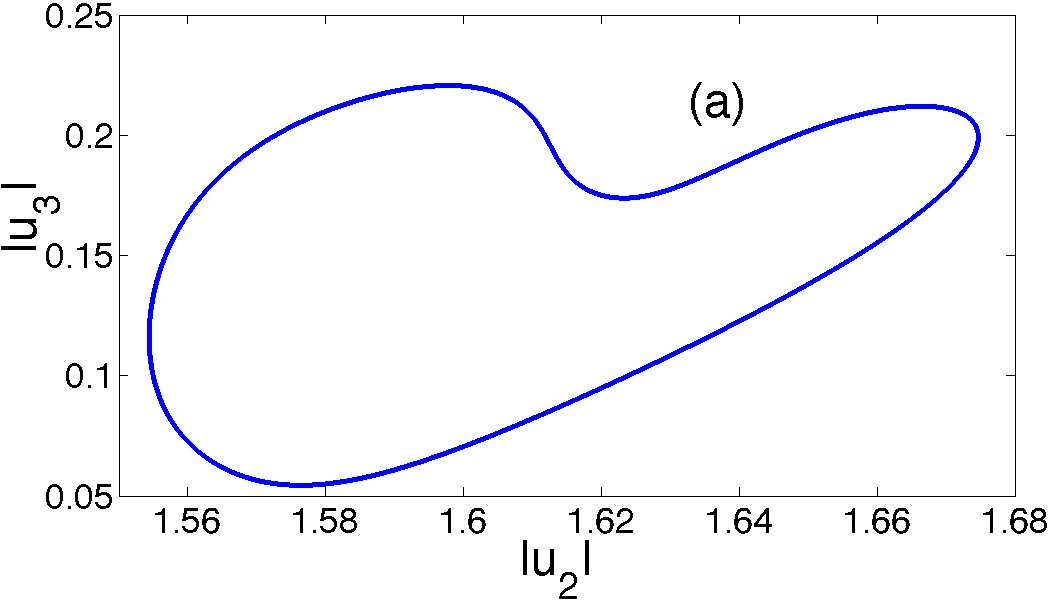}
\includegraphics[width=0.45\linewidth]{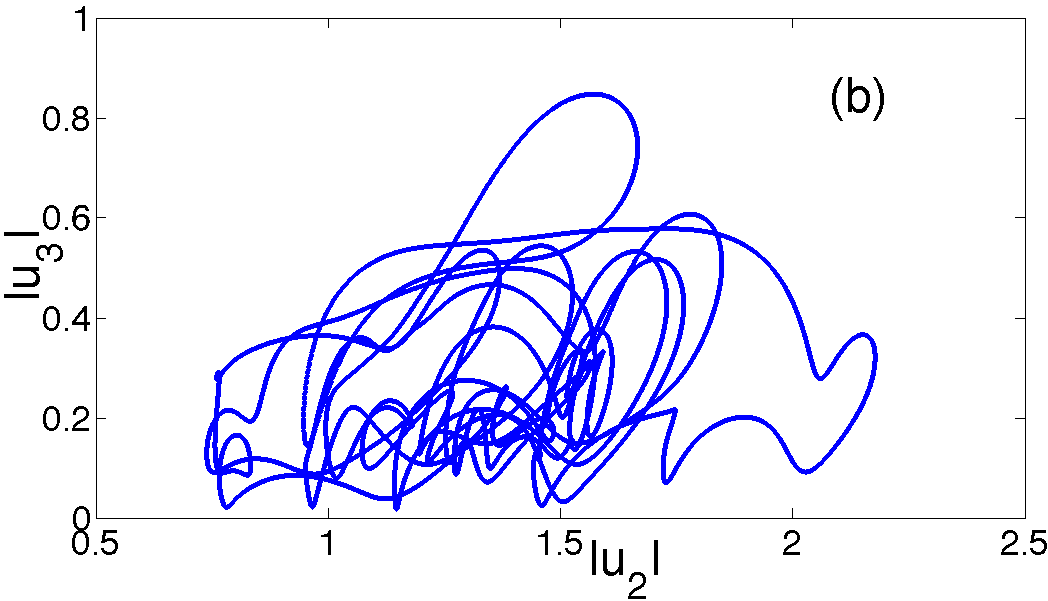}
\caption{The map for $\vert v_{3}\vert$ vs $\vert v_2\vert$, at fixed $Wi = 5$ and (a) $c = 2.0 $ and (b) $c = 10.0$ 
for $\epsilon = 0.3$. 
Like in the case for $\epsilon = 0.5$\ref{fig:map}, we see a similar, clear transition from a non-chaotic to a chaotic behaviour.}
\label{fig:eps0p3_shellmap}
\end{figure*}

A global quantity like the total kinetic energy $K(t)=\sum_n v_n^2(t)$ provides
further insight into the transition to elastic turbulence; its temporal
behaviour with varying Wi and $c$ indeed is an indicator of the changes of
dynamical regime which happen in the system~\cite{BB10}.  In
Fig.~\ref{fig:all_energy} we show time series of $K(t)$ for various
combinations of $c$ and $\mathrm{Wi}$.  For cases with very small values of
$\mathrm{Wi}$---and independent of the value of $c$---the total energy quickly
saturates to an asymptotic value with no noticeable fluctuations, as is typical
for laminar flows (Fig.~\ref{fig:all_energy}, top two panels).  However, as
$\mathrm{Wi}$ increases, even for a small enough value of $c = 1.0$, tiny but
regular oscillations are seen in the temporal dynamics of $K(t)$ vs $t$
(Fig.~\ref{fig:all_energy}, middle panel).  This behaviour is shown clearly in
a zoomed plot in Fig.~\ref{fig:zoomed_energy}(a). Keeping the Weissenberg
number fixed, we now increase the concentration (Fig.~\ref{fig:all_energy},
bottom two panels) and see that the total kinetic energy vs time shows
increasingly chaotic dynamics with large irregular fluctuations. This behaviour
is highlighted in the zoomed-in Fig.~\ref{fig:zoomed_energy}(b).
Figures~\ref{fig:all_energy} and \ref{fig:zoomed_energy} show that the shell
model (which for $c=0$ is laminar because of our choice of parameters), with
increasing effect of polymers characterised by the concentration or the
Weissenberg number, undergoes a transition from a laminar phase to one with
strong fluctuations through a series of intermediate periodic phases for
moderate values of $c$ and $\mathrm{Wi}$.  This phenomenon was first observed
as a function of Wi in direct numerical simulations of the Oldroyd-B
model~\cite{BCAH87} with periodic Kolmogorov forcing~\cite{BB10} and of the
FENE-P model~\cite{BCAH87} in a cellular flow~\cite{G14}.  The shell model
considered here not only reproduces such a transition to chaos through periodic
states as the Weissenberg number is increased, but also shows that an analogous
transition occurs as a function of polymer concentration.  Following
Ref.~\cite{KOJ98}, in Fig. \ref{fig:map} we also show the map $\vert
v_{n+1}\vert$ vs $\vert v_n\vert$, for $n = 2$. The structure of this map for
increasing values of $c$ further shows that the elastic-turbulence regime
emerges through period doubling.

The above results confirm that the shell model with polymer additives
replicates the global features of elastic turbulence. Given the relative
numerical simplicity of shell models, it now behooves us to study in detail the
effects of concentration on the small-scale mixing properties of elastic
turbulence, which determine the importance of this phenomenon for
practical applications.  We quantify mixing in elastic turbulence and its
dependence on $c$ and $\mathrm{Wi}$ by calculating the largest Lyapunov
exponent $\lambda$ of the projection of the shell model on the $v_n$ variables.
We recall that for the fluid GOY shell model such calculations show the
chaotic--non-chaotic transitions as a function of the single parameter
$\epsilon$~\cite{BLLP95,KOJ98}.  For this set of calculations we would like to
ensure that, in the absence of polymers, the flow is non-chaotic in order to
reveal the transition to chaos more clearly. Thus we now study the shell model
with $\epsilon = 0.3$ (see Table~\ref{dec_para}), for which $\lambda = 0$ when
$c = 0$.  In order to check the generality of our conclusions, we use both a
deterministic and a stochastic forcing and find our results insensitive to the
precise nature of forcing. 

Before we turn our attention to a quantitative measure of the transition 
to elastic turbulence below, we immediately note, as seen in Fig~\ref{fig:eps0p3_shellmap}, 
that the basic feature of transition to chaos, with increasing concentration for a 
fixed $\mathrm{Wi}$, persists even for the case of $\epsilon = 0.3$.

\begin{figure}[t!]
\centering
\includegraphics[width=\columnwidth]{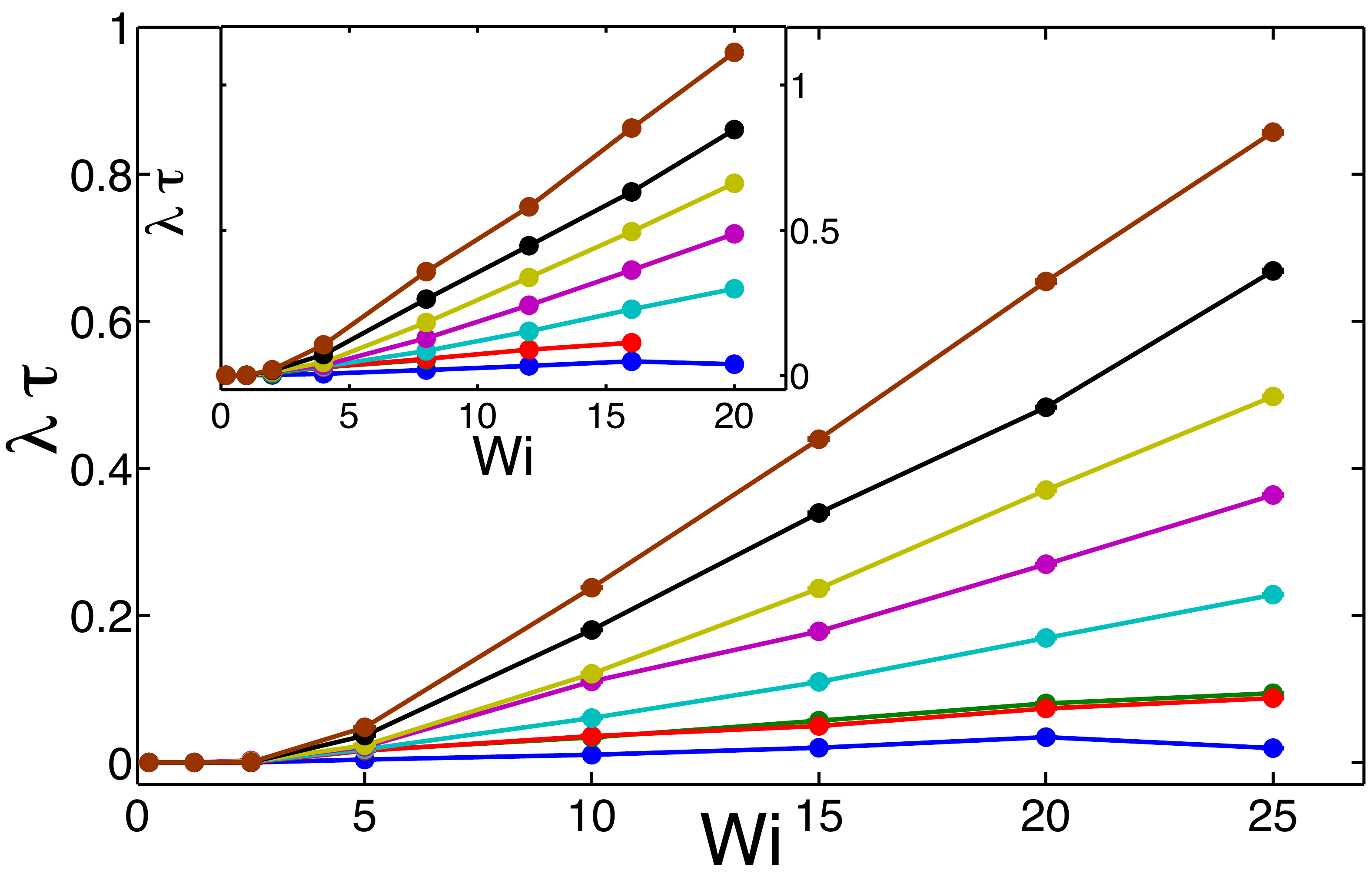}
\caption{$\lambda\tau$ vs the Weissenberg number for $\epsilon = 0.3$ for
various values of the concentration $c$ and for deterministic and stochastic
(inset) forcing. The symbols sizes are proportional to the error bars in our
calculations. The concentration varies between 0 and $20$ from bottom to top.
}
\label{fig:lambda_vs_Wi}
\end{figure}

In Figure~\ref{fig:lambda_vs_Wi} we show the Lyapunov exponent rescaled with
the polymer relaxation time, $\lambda\tau$, as a function of $\mathrm{Wi}$ for
different values of $c$ both for deterministic and for stochastic forcing
(inset).  For small values of $c \lesssim 5$, the rescaled Lyapunov exponent
remains close to 0, and hence the system is non-chaotic or {\it laminar}.  For
sufficiently large values of $c \gtrsim 5$, we find that beyond a threshold
value of the Weissenberg number ($\mathrm{Wi} \approx 5$) the rescaled Lyapunov
exponent increases approximately linearly and for the largest value of $c$, at
$\mathrm{Wi} \gtrsim 25$, we find $\lambda \approx 1/\tau$.  Our findings are
in agreement with analogous calculations made in the viscoleastic Kolmogorov
flow for a single, fixed value of the elasticity of the flow~\cite{BBBCM08}. It
is important to note, though, that the results for the Kolmogorov
flow~\cite{BBBCM08} indicates a slightly more dramatic increase of
$\lambda\tau$ as a function of the Weissenberg number than seen in the shell
model.

\begin{figure}[h!]
\centering
\includegraphics[width=\columnwidth]{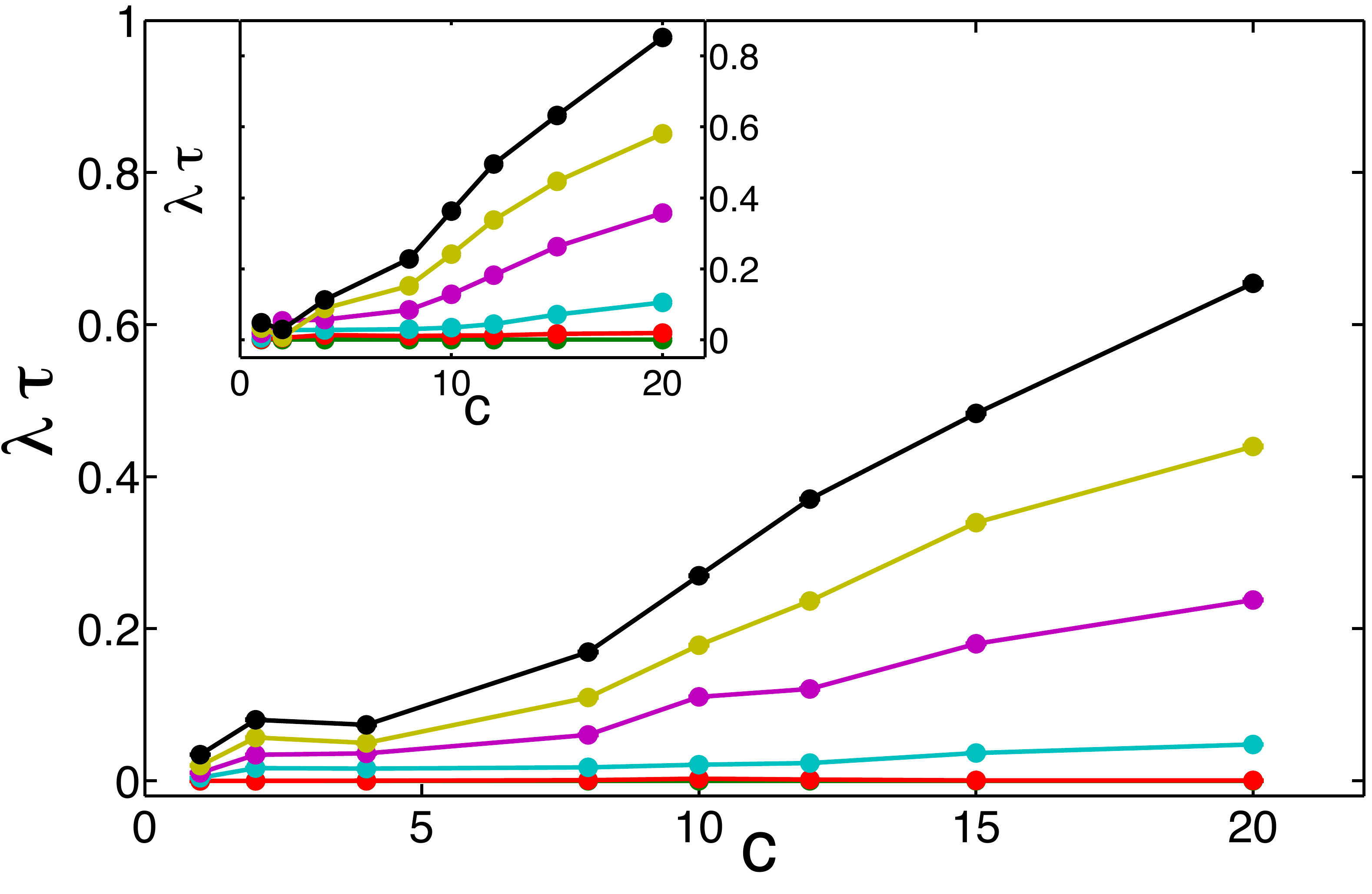}
\caption{$\lambda\tau$ vs the concentration for $\epsilon = 0.3$ for various
values of the Weissenberg number Wi and for deterministic and stochastic
(inset) forcing. The symbol sizes are proportional to the error bars in our
calculations. The Weissenberg number varies from 0 to 25 from bottom to top.
}
\label{fig:lambda_vs_c} 
\end{figure}

We now turn to the behaviour of $\lambda\tau$ as a function of $c$ for
different values of $\mathrm{Wi}$, as shown in Fig.~\ref{fig:lambda_vs_c}.  As
before we find that for low Weissenberg numbers, the flow remains non-chaotic
even when the polymer concentration increases. Beyond a threshold value of
$\mathrm{Wi}$, there is a sharp increase in $\lambda\tau$ when the
concentration becomes greater than 5.  Thus, provided Wi is sufficiently large,
increasing the concentration has a destabilizing effect comparable to that
observed when the Weissenberg number is increased.

\begin{figure}[b!]
\setlength{\unitlength}{\columnwidth}
\includegraphics[width=\columnwidth]{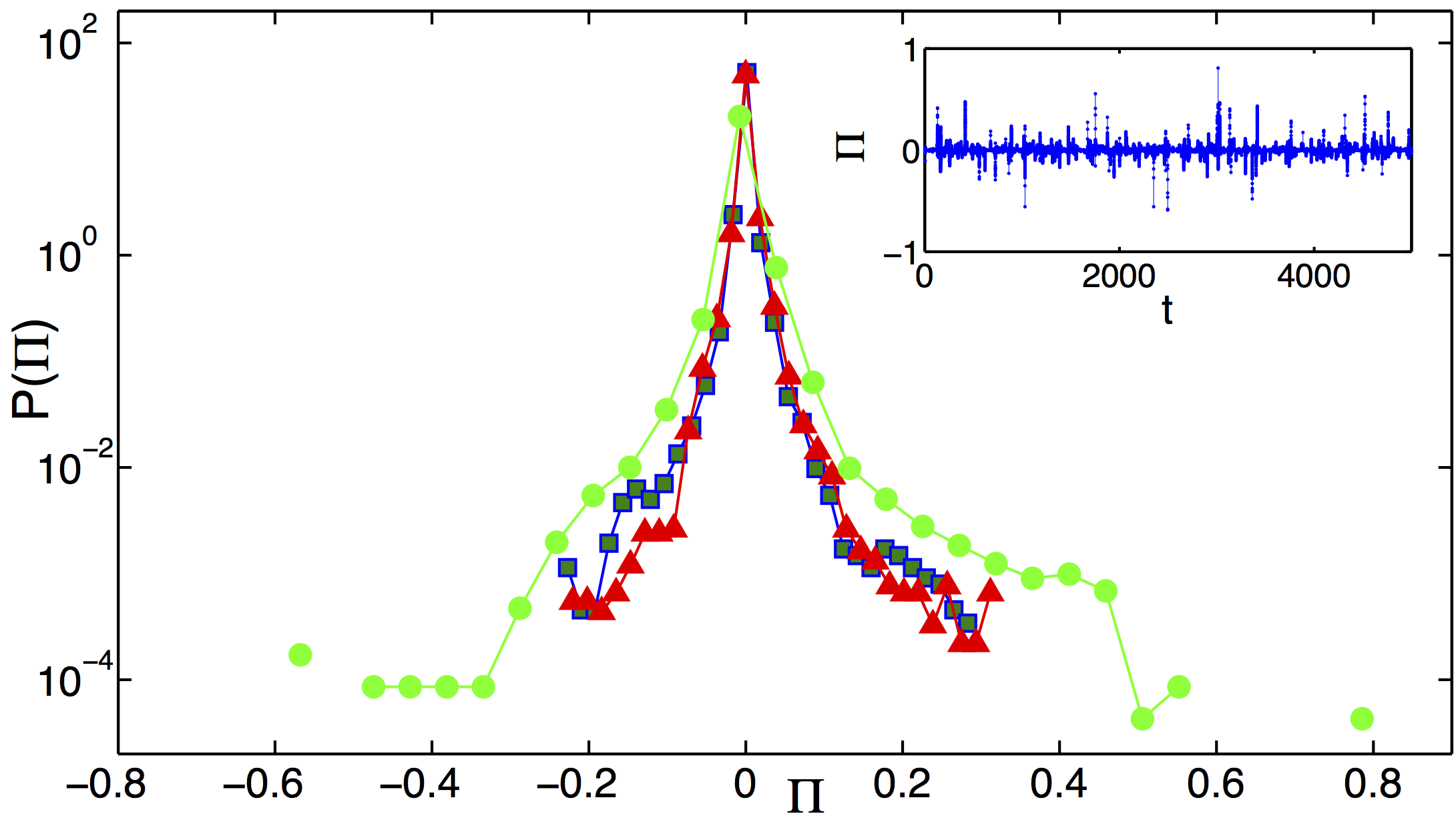}
\caption{Probability density function of $\Pi$ for $\epsilon = 0.3$ at $\mathrm{Wi} = 5$ 
for concentration values $c = 5.0$ (red triangles), 10.0 (blue squares), and 20.0 (green circles). The inset shows a typical 
timeseries of $\Pi$ for $c = 10.0$.}
\label{fig:Pi}
\end{figure}

Finally, the interaction between the polymers and the flow is described by the
energy exchange~\cite{BCP04}: \begin{equation} \Pi=-\dfrac{\nu_p
P(b)}{\tau}\,\mathrm{Re}\Big(\sum_n v_n^*\Phi_{n,bb}\Big).  \end{equation}
Negative values of $\Pi$ indicate that energy flows from the velocity variables
towards the polymers; positive values of $\Pi$ correspond to energy transfers
in the opposite direction. In turbulent drag reduction, the timeseries of 
$\Pi$ is predominantly negative~\cite{BdAGP03,BCP04}.  This fact
signals that polymers drain energy from the flow and justifies the description
of their effect as a scale-dependent effective viscosity~\cite{BCP04}.  In
Fig.~\ref{fig:Pi}, we show the probability density function of $\Pi$ in the
elastic-turbulence regime of the shell model with $\epsilon=0.3$ at different
concentrations for a fixed value $\mathrm{Wi}$.  We find that $\Pi$ takes
positive and negative values with comparable probabilities, i.e. in elastic
turbulence there are continuous energy transfers between the flow and the
polymers without a definite preferential direction (see also inset of 
Fig.~\ref{fig:Pi}.  This result is consistent
with the behaviour of the energy-exchange rate in decaying isotropic turbulence
with polymer additives, the long-time stage of which has properties in common
with elastic turbulence~\cite{WG14}.

\section{Conclusions}

We have considered a shell model of viscoelastic fluid that describes the
coupled dynamics of the velocity and polymer fields in the flow of a polymer
solution. In the regime of large inertia and large elasticity, this model was
previously shown to reproduce the main features of turbulent drag reduction.
We have studied the regime in which inertial nonlinearities are negligible and
have shown that, when the Weissenberg number becomes sufficiently high, the
system shows a transition to a chaotic state.  A detailed analysis of this
chaotic state indicates that the shell model under consideration qualitatively
reproduces the transition to elastic turbulence observed in experiments and in
numerical simulations.  The simplicity of the shell model also allows us to
study the elastic-turbulence regime over a wide range of values not only of the
Weissenberg number but also of polymer concentration.  In particular, we find
that, when the concentration is increased while the Weissenberg number is
fixed, the emergence of the chaotic regime follows a dynamics analogous to that
observed when the Weissenberg number is increased at fixed polymer
concentration. Thus the transition to elastic turbulence shows similar features
as a function of Wi and of~$c$.

This study enhances our understanding of the transition to elastic turbulence
in polymer solutions.  The shell model that we have studied mimicks the
interactions between the Fourier modes of the velocity field and of the polymer
end-to-end separation field in a viscoelastic fluid, but it contains no
information on the spatial structure of these fields.  The fact that such a
model can replicate the main features of elastic turbulence shows that the
specific geometrical configuration of the system does not play an essential
role in the transition to elastic turbulence and that the physical mechanisms
leading to elastic turbulence do not rely on the boundary conditions or on the
mean flow.

\section{Acknowledgments}

The authors are grateful to \textsc{Chirag Kalelkar} and \textsc{Stefano
Musacchio} for useful discussions. This work was supported in part by the
Indo--French Centre for Applied Mathematics (IFCAM) and by the EU COST Action
MP 1305 ``Flowing Matter''.  SSR acknowledges support from the AIRBUS Group
Corporate Foundation in Mathematics of Complex Systems established at ICTS and
the hospitality of the Observatoire de la C\^ote d'Azur, Nice, France. DV 
acknowledges the hospitality of ICTS-TIFR, Bangalore, India.

\end{document}